\documentstyle[11pt,aaspp4]{article}

\def \hb {H$\beta$}
\def \hg {H$\gamma$}
\def \hd {H$\delta$}
\def \oiii  {{\rm [O~III]}}
\def \oii   {{\rm [O~II]}}

\def \nii   {{\rm [N~II]}}
\def \sii   {{\rm [S~II]}}
\def \nev   {{\rm [Ne~V]}}

\def \fevii {{\rm [Fe~VII]}}
\def \ariv  {{\rm [Ar~IV]}}
\def \mgii  {{\rm Mg~II}}
\def\wave#1{$\lambda${#1}}
\def\waves#1{$\lambda\lambda${#1}}
\def\etal{{\rm et al.}}

\def\arcsp{\hbox to 1pt{}\rlap{\arcsec}.\hbox to 2pt{}}
\def\ltsima{$\; \buildrel < \over \sim \;$}
\def\simlt{\lower.5ex\hbox{\ltsima}}
\def\gtsima{$\; \buildrel > \over \sim \;$}
\def\simgt{\lower.5ex\hbox{\gtsima}}

\def   \flam    {$F_\lambda$}
\def   \pf      {$P\times F_\lambda$}
\def   \qf      {$Q\times F_\lambda$}
\def   \uf      {$U\times F_\lambda$}
\def   \kms     {km s$^{-1}$}
\def   \cc      {cm$^{-3}$}
\def   \p       {$P$}

\def   \pa      {$\theta$}

\def \fluxu {ergs cm$^{-2}$ s$^{-1}$ \AA$^{-1}$} 
\def \lsun      {$L_{\sun}$}

\def \hubu      {km s$^{-1}$ Mpc$^{-1}$}
\def \ebv       {$E(B-V)$}
\def \objn      {IRAS P09104+4109}
\def \magp      {$A_{\rm{M/I}}$}

\begin{document}

\title{Keck Observations of the Hidden Quasar IRAS P09104+4109}
\author{Hien D. Tran\altaffilmark{1}, 
Marshall H. Cohen\altaffilmark{2},
and
Montse Villar-Martin\altaffilmark{3, 4}
}
\altaffiltext{1}{Department of Physics \& Astronomy, Johns Hopkins University,
Baltimore, MD 21218; tran@pha.jhu.edu.} 
\altaffiltext{2}{Astronomy Department, California Institute of Technology, 
Pasadena, CA 91125; mhc@astro.caltech.edu.}
\altaffiltext{3}{Institut d'Astrophysique de Paris, F75014 Paris, France.}
\altaffiltext{4}{Current address: Department of Physical Sciences, University of Hertfordshire, Hatfield AL109AB, UK; mvm@star.herts.ac.uk.}

\begin{abstract}

We present imaging and spectro- polarimetric observations of the
ultraluminous infrared galaxy IRAS P09104+4109 using the Keck 10-m Telescope.
We detect the clear presence of broad \hb, \hg, and Mg~II~\wave 2800 emission 
lines in the polarized flux spectra of the nucleus and of an extranuclear 
emission region $\sim$ 4\arcsec~away, confirming the presence of a hidden 
central quasar. 
The polarization of the broad Mg II emission line is high ($\sim$ 29\%), consistent with
the remarkably high polarization ($\sim$ 30\%--40\%) observed in the extended 
continuum emission. This indicates that the off-nuclear continuum is dominated by 
light scattered from the hidden quasar, most probably by dust mixed with the 
line emitting gas. 
The high polarizations, combined with the ``foreshortened'' morphology of the
polarized brightness distribution allow us to constrain
the scattering biconical structure to be at inclination $i~\approx~50$\arcdeg~with a 
half-opening cone angle $\theta_c~\approx$~40\arcdeg.

The narrow emission lines are polarized in a stratified fashion, with the high 
ionization lines (\oiii, \nev, \fevii) being polarized 0.7\%--1.7\% and 
\oii~essentially unpolarized. 
The line polarizations are positively correlated with critical density,
ionization potential, and velocity width of the emission lines.
This indicates that, as is the case with the narrow-line radio 
galaxies, which also often contain powerful quasars, the 
narrow-emission line region may be partially shadowed by the putative torus, 
with the higher ionization lines originating closer to the nucleus.  
 
One notable characteristic of the extranuclear knot is that all species of 
Fe are markedly absent in its spectrum, while they appear prominently in the 
nucleus.
In addition, narrow Mg II is observed to be much weaker than predicted by 
ionization models.
Our favored interpretation is that there is a large amount of dust
in the extranuclear regions, allowing gaseous refractory metals to deposit. 
Near the nucleus, dust is destroyed in the strong radiation field of the 
quasar, inhibiting metal depletion onto grains.
The extended emission regions are most likely material 
shredded from nearby cluster members and not gas condensed from the cooling 
flow or expelled from the obscured quasar. 

The higher temperature inferred from [O III] lines compared to that from 
[N II] and the general better agreement with models of line ratios, 
especially \oiii~\wave 5007/\wave 4363 and He~II/\hb, 
provide strong evidence for matter-bounded 
clouds in addition to ionization-bounded clouds in the NLR. 
Ionization by pure velocity shocks can be ruled out. Shocks with 
photoionizing precursors may be present, but are probably not a dominant 
contributor to the energy input.

\end{abstract}

\keywords{galaxies: active --- galaxies: individual (\objn) --- galaxies: Seyferts --- infrared: galaxies --- polarization}

\section{Introduction}

Discovered by the $IRAS$ all-sky survey,
ultraluminous infrared galaxies (ULIRGs, see review by \cite{sm96}) 
emit most of their energy in the infrared ($L_{IR} > 10^{12}$ \lsun), and 
could harbor infant quasars enshrouded in a large amount of dust 
(\cite{san88}). On the other hand, they 
may also represent energetic, compact starbursts (\cite{con91}; \cite{gen98}). 
Much recent research effort has been devoted to 
understanding the dominant energy source in these 
ULIRGs -- whether it is obscured quasars or intense bursts of star formation. 

IRAS P09104+4109 ($z$ = 0.44; \cite{k88}) is one of the few
exceptionally luminous ULIRGs sometimes referred 
to as the ``hyperluminous'' ($L_{IR}~\simgt~10^{13}$~\lsun) infrared galaxies.
The others include F15307+3252 ($z$ = 0.926; \cite{c94}) and F10214+4724 
($z$ = 2.286; \cite{rr91}). 
All of these galaxies have been shown to harbor a quasar nucleus
obscured from direct view (P09104+4109, \cite{hin99};
F15307+3252, \cite{hin95}; F10214+4724, \cite{goo96}). 
This has led to the suggestion that perhaps all warm\footnote{``Warm'' ULIRGs 
are those having $f_{25}/f_{60}~>~0.2$, \cite{low88}; \cite{san88});
$f_{25}$ and $f_{60}$ are the $IRAS$ flux densities in units of
Jy at 25 $\mu$m and 60 $\mu$m, respectively.}
ULIRGs contain buried QSOs and that they may be the misdirected type 2 QSOs 
(AGN with Seyfert 2-like emission-line characteristics but QSO-like 
luminosities).
A diagnostic diagram (\cite{t99}) involving \oiii~emission line and 
infrared color $f_{25}/f_{60}$ does indeed show that essentially all warm 
ULIRGs with sufficiently high ionization harbor energetic quasars in their 
centers.

IRAS P09104+4109 has been identified with a central cD galaxy in a rich cluster
(\cite{k88}; Hall \etal~1997).
It has a high-ionization spectrum characteristic of a Seyfert 2 galaxy.
Ground-based and $Hubble~Space~Telescope$ ($HST$) images show that 
\objn~has an off-nuclear northern extension or ``plume'' of ionized gas 
(\cite{k88}; \cite{hn88}; \cite{armus99}) that 
may serve as a scattering mirror of the light originating from the nucleus.
We wish to probe the origin of this extension and constrain the scattering 
geometry for \objn~by obtaining polarimetric observations of this knot.

In this paper, we confirm the high polarization and the broad emission 
lines of \hb, \hg~and Mg II in polarized flux of the nucleus. 
In addition to the $B$-band imaging polarimetry, we also present new 
spectropolarimetric data for the nucleus and off-nuclear emission 
regions of \objn. 
We study in detail the emission-line spectra of the nuclear and extra-nuclear 
regions in order to derive an understanding of the ionization mechanism 
of the narrow-line region (NLR) of this galaxy.
Throughout this paper, we assume $H_o$ = 75 \hubu, $q_o$ = 0 and $\Lambda$ = 0.
At the redshift $z=0.44$ of P09104+4109, 1\arcsec~corresponds to a projected
size of 5 kpc.

\section{Observations and Reductions}

Spectropolarimetric observations were made with the polarimeter 
(\cite{coh97}) installed in the low resolution imaging spectrometer 
(LRIS, \cite{oke95}) on the 10-m Keck I telescope on the night 
of 1994 December 31 (UT).
A 1\arcsec~wide, long slit, was centered on the nucleus of \objn,
and oriented at $PA$ = 14\arcdeg~to include the extended 
\oiii~structure. 
The emission-line regions extend $\sim$ 5\arcsec~north-northeast of the nucleus 
(\cite{armus99}), as well as a few arcsecs south. 
We used a 300 grooves 
mm$^{-1}$ grating which provided a dispersion of 2.46~\AA~pixel$^{-1}$ 
and a resolution of $\sim$~10~\AA~(FWHM).
The observations were made by following standard procedures of rotating 
the half waveplate to four position angles (0\arcdeg, 22.5\arcdeg, 45\arcdeg, 
and 67.5\arcdeg), and dividing the exposure times equally among them. 
Two sets of observations, lasting 60 and 52 min., were made under photometric 
conditions, giving a total exposure time of 112 min. The results presented
here are the average of the two sets.

Spectra of the nucleus were extracted from an aperture 3\arcsp4 wide, 
and those of the NE off-nuclear region were extracted from a region 
3\arcsp6 wide centered 3\arcsp4 from the nucleus. 
This includes the entire NE plume seen in the HST images by 
Armus \etal~(1999).
The data were reduced using VISTA, following standard polarimetric reduction 
techniques as described by Cohen \etal~(1997).
Hereafter, we shall refer to the nucleus spectrum as NUC and that of the NE 
extension as EXT.

Imaging polarimetry was obtained on the night of 1995 December 16 with the 
Keck I telescope.
The images were taken through the $B$ filter, which contains mainly continuum
emission. Exposure times were 10 min. at each of the four waveplate positions.
Sky was clear but seeing was variable between 1\arcsec -- 1\arcsp8.
Reductions were done analogously to spectroscopic observations as described
by Cohen \etal~(1997).

\section{Results}
\subsection{Emission Line Ratios}

\begin{figure}
\plotone{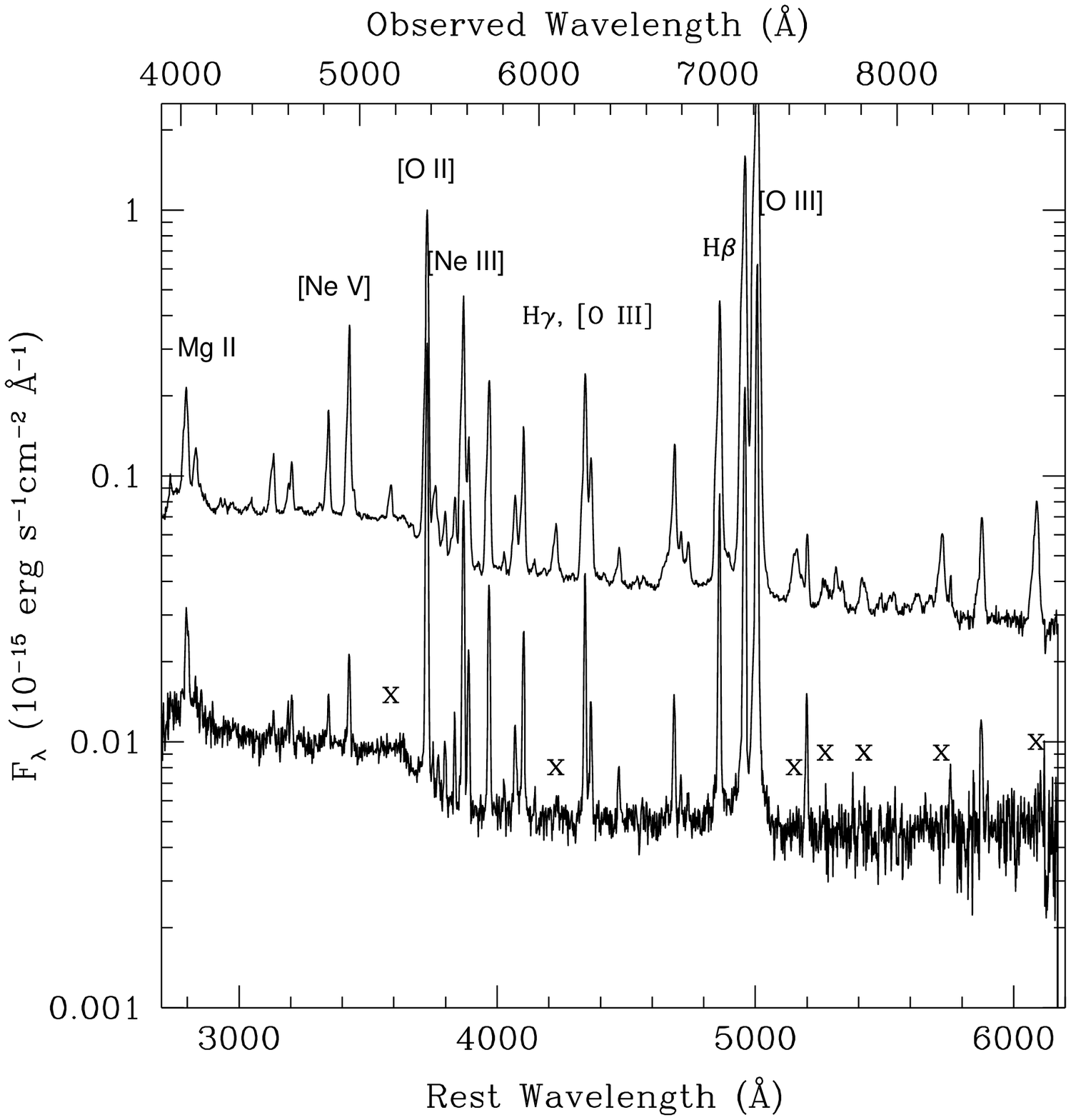}
\caption{Comparison of the total flux spectra of the nucleus ({\it top}) and
off-nuclear NE extension region ({\it bottom}). 
``X'' denotes where Fe line is expected but not seen in the extension spectrum. 
Note the complete disappearance of Fe lines, in order from left to right: 
[Fe VII] \wave 3588, [Fe V] \wave 4229,  [Fe VI+VII] \waves 4146, 5159, 
[Fe III+VII] \waves 5270, 5276, [Fe VI] \waves 5424, 5427,
[Fe VII] \wave 5720, [Fe VII] \wave 6087.
\label{nucextsp}}
\end{figure}

Table 1 presents the integrated emission-line flux ratios with respect to
\hb~and their rest-wavelength equivalent widths for the NUC and EXT spectra, 
shown in Figure \ref{nucextsp}. 
The emission line fluxes of NUC were measured from the starlight-subtracted 
spectrum. The contribution from an old 
stellar population represented by the elliptical galaxy NGC 821 was 
used to remove the starlight. 
The starlight fraction ranges from 5\% in the blue end to a maximum of 40\%
in the red. We did not correct for starlight in the EXT spectrum as the 
galaxy contribution in the outer regions is negligible.
The largest source of uncertainty in the flux measurement is the placement of
the continuum. For strong lines (F/F(\hb) \simgt 20\%) the uncertainty is \simlt 10\%;
for weaker lines (F/F(\hb) \simlt 20\%) the uncertainty is 10\%--25\%.

The line ratios in NUC have been corrected for a small amount of extinction 
\ebv = 0.24 determined from the \hg/\hb, and \hd/\hb~ratios, assuming
the intrinsic case B values of 0.47 and 0.26, respectively (Osterbrock 1989),
and the extinction curve of Cardelli, Clayton, \& Mathis (1989) with $R_V=3.1$.
This correction is small and does not significantly alter the 
conclusions of this paper. 
No corrections have been applied to EXT as its observed Balmer line ratios 
are close to case B values, indicating little extinction due to foreground dust. 

The small extinction derived might seem surprising, given the high infrared luminosity
of this object. However, this is consistent with the result reported by 
Veilleux, Kim \& Sanders (1999) in their spectroscopic study of a large sample of ULIRGs, 
that there is no obvious correlation between extinction and
IR luminosity among the IR-bright galaxies.
The mean \ebv~for the Seyfert 2 galaxies in 
their sample is 1.2 with a range of 0.3 to 2.7.
Our reddening estimate is intermediate between those found by Kleinmann \etal~(1988, \ebv~$<$ 0.12), 
who used the same optical line ratios, and Soifer \etal~(1996, \ebv~= 0.38--0.79), who 
used infrared [S II] and hydrogen lines in combination with optical lines.
Soifer et al. also suggested that the low reddening may result from the dust being
confined in a disk-like structure, which effectively obscures the central nucleus but does 
not block our line of sight to the NLR. As Veilleux \etal~(1999) noted, the color excess 
derived from the optical line ratio method tends to underestimate the amount of dust in 
these objects.

In the nucleus, the narrow lines all display a strong blueward asymmetry, 
as reported by Kleinmann \etal~(1988), Crawford \& Vanderriest (1996) and Hines \etal~(1999),
and cannot be adequately fit with a single Gaussian profile. 
Two Gaussian profiles including 
a main, central component with (intrinsic) FWHM $\approx$ 600 \kms~and 
another component blueshifted 720--980 \kms~with FWHM $\approx$ 700 \kms~are 
able to fit the observed profiles well.
Our measured range of velocity shifts of the blue-shifted component
is somewhat smaller than the 1250 \kms~reported by 
Crawford \& Vanderriest (1996), and more consistent with 950 \kms~reported by
Kleinmann \etal~(1988).

Aside from the smaller line widths in EXT, the most remarkable 
difference between the NUC and EXT spectra is the complete absence of any
Fe species in EXT, as Figure \ref{nucextsp} shows. 

\subsection{Imaging Polarimetry}

\begin{figure}
\includegraphics{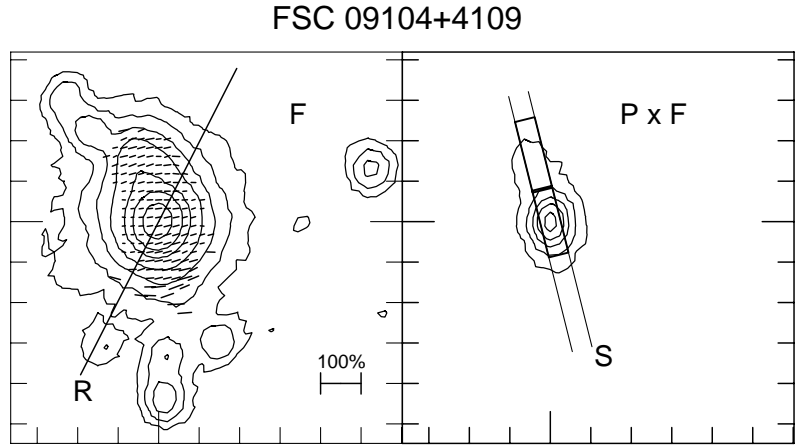}
\vspace{3in}
\caption{Polarization map for \objn~obtained in $B$-band with the
Keck telescope. The left panel shows the polarization vectors superimposed 
on the contour plot of the total flux, with north up and east to the
left. The scale is 2\arcsec~per tick, and the contours are 0.5, 1, 2, 4,
10, 20, and 50\% of the peak. 
Polarization vectors are shown for those pixels where $P > 2.5\sigma$.
The line marked ``R'' shows the radio axis. A horizontal bar representing 
100\% polarization is shown at the bottom right. 
The right panel shows the polarized flux $P\times F$. The contours are
5, 15, 30, 50, and 80\% of the peak. The polarized light distribution peaks
at the flux peak, suggesting that the observed light 
is dominated by scattered radiation. The 1\arcsec~wide spectroscopy slit (``S'') is 
indicated, along with the extraction apertures for the NUC and EXT spectra.
\label{polmap}}
\end{figure}

The imaging polarization of \objn~is shown in Figure \ref{polmap}. 
The polarization is very high ($\sim$ 25\%--30\%) throughout the nucleus and 
extensions up to 5\arcsec~away. 
The polarization increases strongly across the source. 
In the south, a number of
pixels shows $P >$ 30\%, and the highest is 40.3\% $\pm$ 12.8\%. 
We note that the distribution of surviving vectors is biased because those 
which have positive errors are preferentially saved; however, there is no
doubt that the polarization is remarkably high.   

Unlike most of the narrow-line radio galaxies like 3C 195 and 3C 33
(\cite{coh99}), there is no well-defined V-shape bi-cone of scattered light, 
although the entire region is highly polarized.
The polarization field displays a strong similarity to
the patterns in the BLRG FSC 2217+259 and the highly-polarized NLRG 3C 234
(\cite{t95}; \cite{coh99}).  
To the NE and SW, the vectors are perpendicular to the radius and this suggests
scattering from a central illuminating source. 
However, the curvature of the polarization vectors is small, as in 3C 234 and FSC 2217+259.  
This may suggest that the viewing angle is small, indicating that the line of sight is  
just outside the ionization cone (see \S 4.3), a geometry that has also been suspected in 
3C 234 (\cite{tcg95}; \cite{coh99}). 

The southern extension appears to display somewhat higher polarization than
its northern counterpart. This may be a result of forward scattering and would
imply that the cone axis is pointed toward us in the south, analogous to 
Cygnus A (\cite{ogl97}).
Dilution by line emission is not a major factor since there are no strong
emission lines through the $B$ band used. However, dilution by hot stars or 
nebular continuum could play an important role in the north as there is
more ionized gas there.  
Our imaging polarimetry confirms and extends the polarization map obtained
with $HST$ by Hines \etal~(1999), which shows a reflection ``cone'' in the 
central 1\arcsec. 

The right panel of Figure \ref{polmap} shows the polarized flux $P \times F$ which is
commonly called the ``polarized light''.
As can be seen, the polarized light distribution extends more to the north than to
the south, reaching up to $\sim$ 4\arcsec~(20 kpc) to the northeast along $PA$ = 19\arcdeg. 
This coincides well with the extended [O III] emission seen in 
the ground-based (\cite{k88}) and $HST$ images (\cite{armus99}).
The polarized extension is thus better aligned along this emission-line structure 
($PA \approx$ 11\arcdeg) than the radio axis ($PA$ = 333\arcdeg), a condition also 
exhibited by many high and low-redshift radio galaxies (\cite{t98}; \cite{coh99}).
This indicates that EXT serves as a reflection
region of nuclear light, and that it lies exposed to the ionizing radiation 
escaping through the ionization cones from the hidden quasar.  
The polarized flux distribution peaks at the total flux peak, suggesting that scattered
light dominates the observed radiation. 

\subsection{Spectropolarimetry}

\subsubsection{Nucleus}

\begin{figure}
\plotone{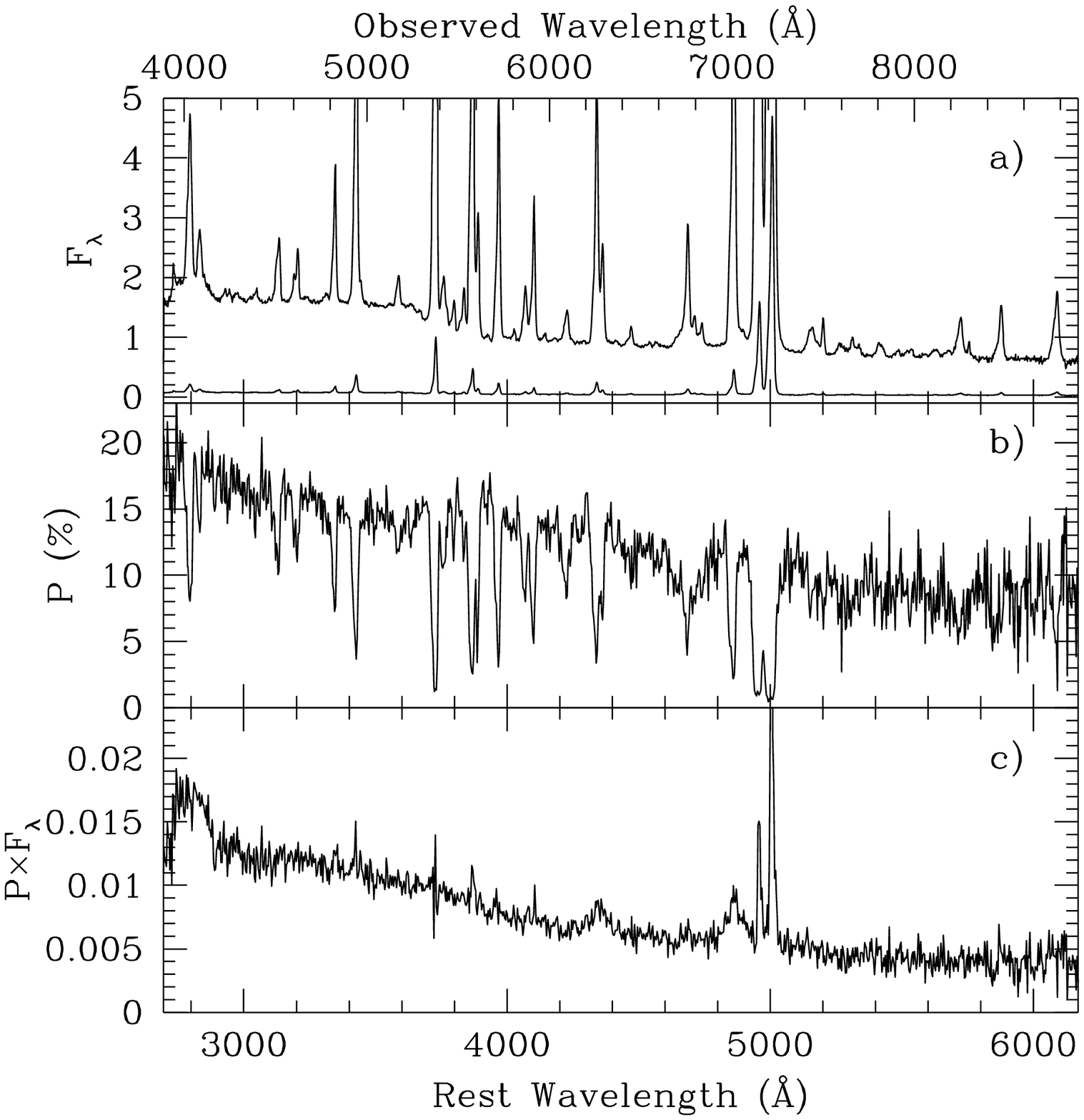}
\caption{Spectropolarimetry of the nucleus of \objn. 
{\it (a)} Total flux spectrum \flam, displayed at two scales to show the 
weaker emission lines, 
{\it (b)} observed degree of polarization \p, and 
{\it (c)} polarized flux spectrum \pf.
The flux scales are in units of 10$^{-15}$ \fluxu.
Broad \hb, \hg, and Mg II, as well as narrow lines of \oiii, \nev, and 
[Ne III], are clearly present in the \pf~spectrum. 
\label{nucspol}}
\end{figure}

The spectropolarimetry for NUC is displayed in Figures \ref{nucspol}. 
High polarizations are seen, with magnitudes consistent with those seen in the imaging 
polarimetry (Fig. \ref{polmap}).
The Keck data of the nucleus confirm the general results of Hines \etal~(1993, 1999).
The observed continuum polarization uncorrected for starlight ranges 
from $\sim$ 8\% at 6000 \AA~(rest) to about 18\% at 3000 \AA (rest).
The polarization $PA$ is constant with wavelength at 97\arcdeg.
In addition to broad \hb, \hg, and probable Fe II multiplets near 
3200 \AA~reported by Hines \etal, we also detect broad Mg II in polarized flux.
We measured a FWHM of 7540 $\pm$ 700 \kms~for the broad \hb~in polarized
flux, somewhat smaller than that reported by Hines \etal~(1999, 12,000 $\pm$ 2500\kms). 
This is comparable to the FWHM of 6000 $\pm$ 700 \kms~measured for \hg.
The resemblance of the nuclear \pf~to a quasar spectrum also extends to its 
shape. 

 
The galaxy-corrected total flux spectrum \flam~($\alpha=-0.94$, 
$f_\nu \propto \nu^\alpha$) is 
redder than \pf~($\alpha=-0.46$), which is similar in slope to 
the more core-dominated composite QSO spectrum of Baker \& Hunstead 
(1995, $\alpha=-0.5$), or that of Francis \etal~(1991, $\alpha=-0.32$).
Our measured spectral indices agree well with those of Hines \etal~(1999).
This suggests that there is a second ``featureless continuum'' FC2 in the 
total flux spectrum that makes it redder.  
This component is likely due to a combination of hot stars and nebular 
continuum plus reddened transmitted light from the hidden AGN 
(e.g., \cite{coh99}). 
Quasars viewed at high inclination angle have been shown to exhibit redder 
continuum due to dust extinction (\cite{bak97}).
The presence of hot stars, perhaps arising in a starburst, 
is suggested by the broad feature underlying He II \wave 4686. 
This feature cannot be a broad scattered He~II component since it appears very 
strongly in the total flux spectrum, suggesting that the light is viewed 
directly, and it is not detected in the \pf~spectrum (see Fig. \ref{nucspol}). 
It can be attributed to Wolf-Rayet (W-R) stars, as has been 
identified by Heckman \etal~(1997) in Mrk 477 and Storchi-Bergmann, Cid 
Fernandes, \& Schmitt (1998) in Mrk 1210, and Tran \etal~(1999) in 
TF~J1736+1122.

In addition, there is nebular continuum giving rise to 
the Balmer jump near 3650\AA. 
If this is the case we would expect the Balmer jump to be unpolarized. Is it?
To answer this, we have constructed an 
``unpolarized spectrum'' = \flam $-(1/0.20)$ $\cdot$ \pf,
assuming an intrinsic continuum polarization of 20\%. 
We then measured the Balmer
jump index defined as $f_{3727-}/f_{3727+}$, where $f_{3727-}$ is the continuum
flux just blueward of \oii~\wave 3727, and $f_{3727+}$ that just redward 
of it. We obtain an index of 1.74 $\pm$ 0.1 for the total flux spectrum,
and 2.0 $\pm$ 1 for the unpolarized flux. 
Naturally, there is a larger uncertainty in the value obtained from the
unpolarized flux spectrum. However, a Balmer discontinuity appears to still be
present, meaning that it is unpolarized, and supporting the notion that FC2 
contains nebular continuum plus starlight from a starburst.

We find that $P_s$, the intrinsic polarization\footnote{We refer to the 
continuum-subtracted polarization of the broad emission line as the 
``intrinsic polarization''.} produced by scattering, is 
$P_s$(\hb) $\approx$ 16\% while $P_s$(Mg II) $\approx$ 29\% (see below).
This strengthens the reality of the nebular continuum discussed above, 
since a rising $P_s$ toward the blue would leave 
most of the Balmer discontinuity in the unpolarized flux spectrum.
This blue $P_s$ is possible with dust but not electron scattering. 
Thus, dust scattering is important, while electron scattering could 
contribute but not dominantly.
 
\subsubsection{Extension Polarization Spectrum}
 
\begin{figure}
\plotone{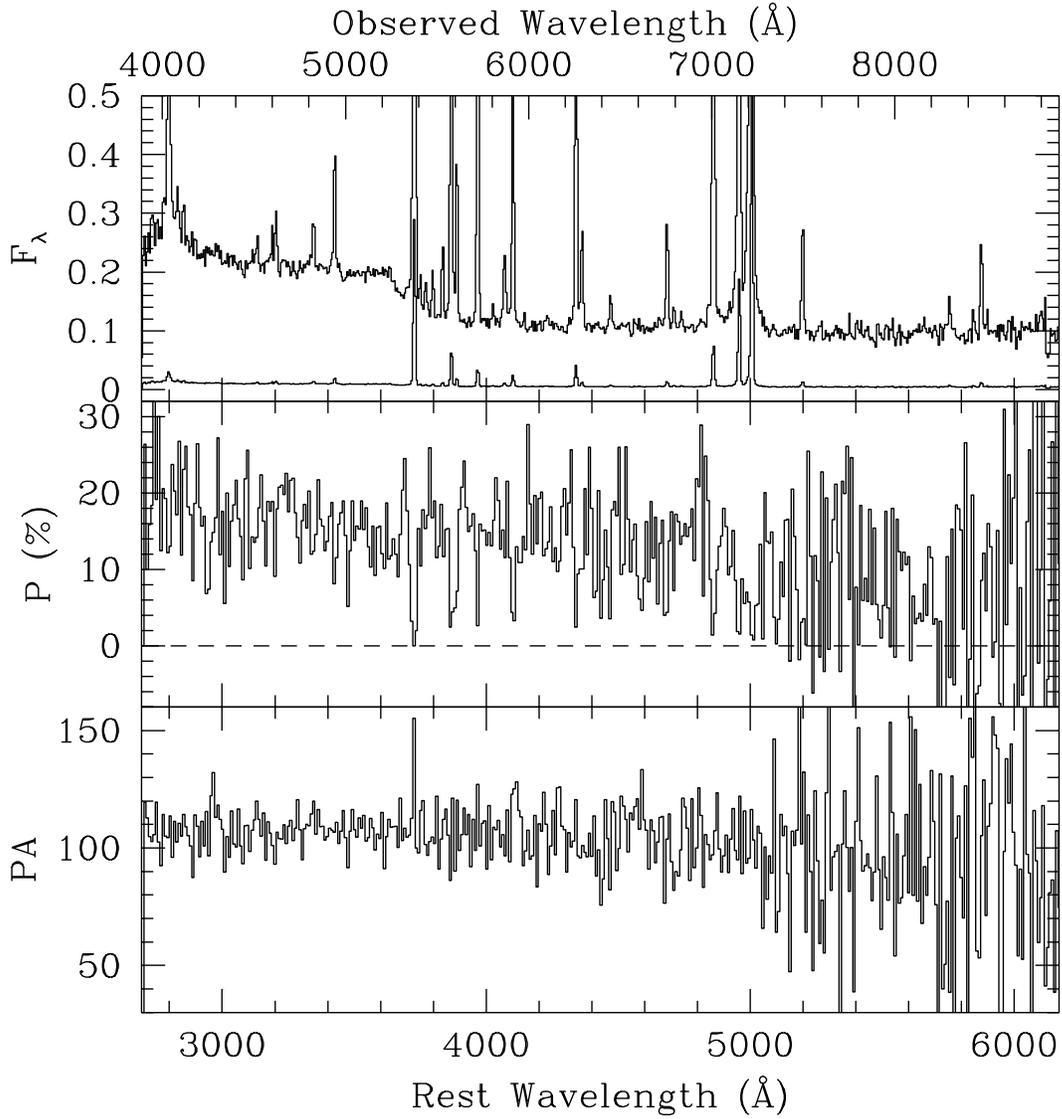}
\caption{Spectropolarimetry of the NE off-nuclear emission line region 
of \objn. From top to bottom panels are: the total flux, \p, and 
polarization $PA$. \p~and $PA$ have been binned 5 pixels and the flux has 
been binned 2 pixels. The polarization shows similar high magnitude and 
wavelength dependence as in the nucleus. 
\label{extspol}}
\end{figure}

The polarization of EXT is noisier than NUC, and we have binned the results
(except for \flam), presented in Figure \ref{extspol}, by 5 pixels to 
improve S/N. 
As can be seen, the polarization shows similar behavior and magnitude 
to those in NUC.
In addition, a broad Mg II is clearly seen in the total flux spectrum. 
This indicates that the continuum in EXT is dominated by scattered light 
from the obscured quasar. 
The polarization $PA$ is roughly constant with wavelength at 
106\arcdeg~$\pm$ 2\arcdeg, slightly higher than that measured in the nucleus.
The $PA$ is accurately perpendicular to the slit, as expected for an off-nuclear
measurement of a reflection nebula (see \cite{coh99} for similar cases).

\begin{figure}
\plotone{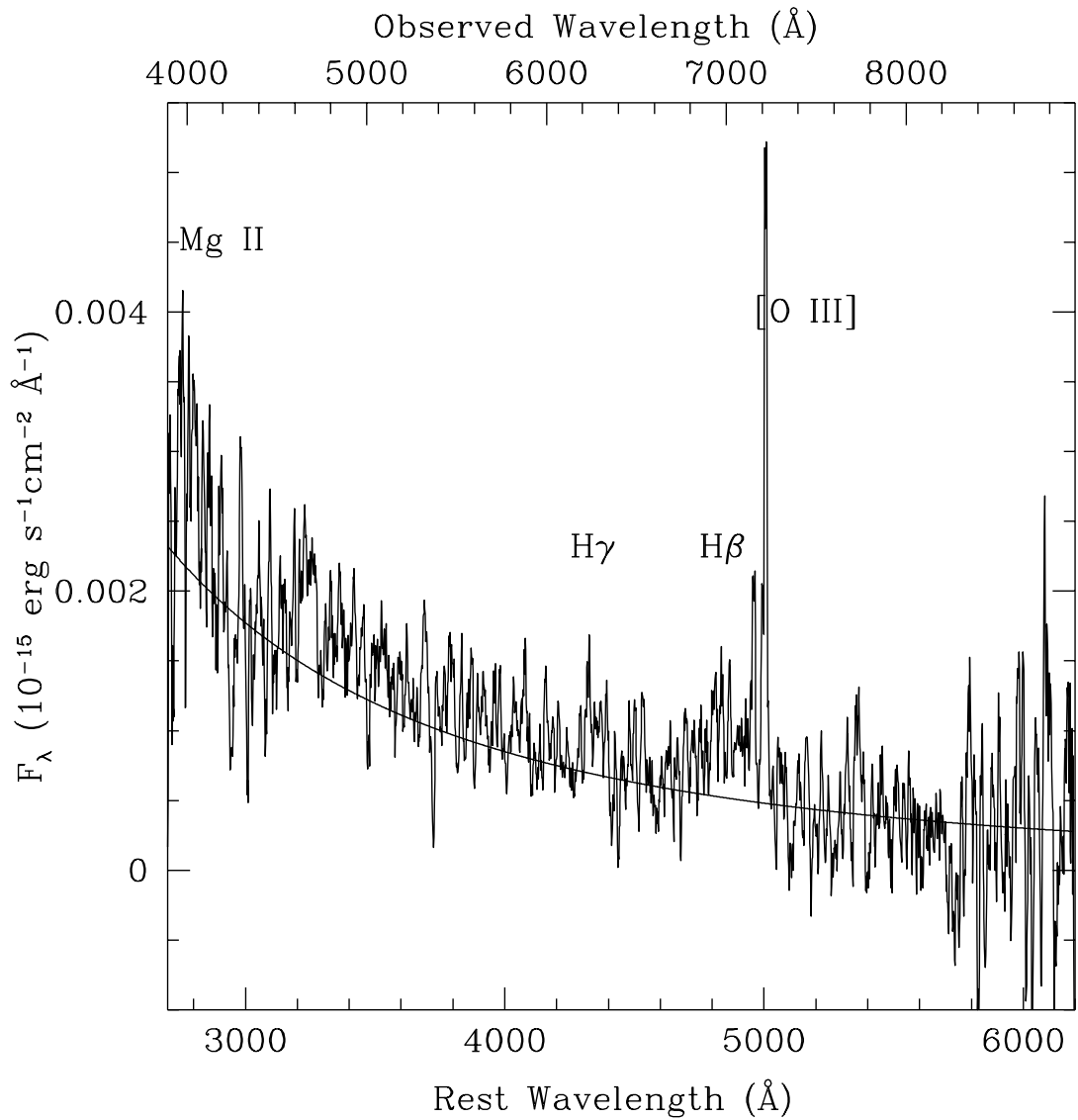}
\caption{The polarized flux spectrum \pf~of the extension, smoothed 
7 pixels to improve S/N. A broad \hb, \mgii~and perhaps \hg~are visible
atop a blue continuum with $\alpha = +0.5$ (solid line, $f_\nu \propto \nu^\alpha$).
\label{extpf}}
\end{figure}

We show the EXT \pf~smoothed by 7 pixels in Figure \ref{extpf}, which 
shows clearly that broad \hb~and Mg II are present.
In \pf, the spectral index is $\alpha=+0.5$, significantly bluer 
than the average quasar spectrum. 
Assuming that the incident spectrum is typical for a quasar, this suggests that, 
as for the nucleus component, dust is the dominant 
scatterers which could bluen the light. 

\subsubsection{Modeling of the Broad Line Polarization}

We model the spectropolarimetric observations using the procedure 
described by Tran \etal~(1997), to separate the various components of the 
emission lines and continuum.
In a region surrounding the \oiii+\hb~complex,
we perform a fit of the continuum and emission-line components in the 
total flux, \qf, and \uf~spectra. 
The broad \hb~is assumed to have the same profile as in the polarized flux 
spectrum. The narrow lines of \hb~and \oiii~were fitted assuming 
that they are similarly polarized, and consist of one main core component 
and another blue-shifted component, both with Gaussian profiles. 
Our Keck data do not have sufficient spectral resolution to warrant 
the addition of more than two components in the \oiii~profile, and we do
not attempt to fit the line polarizations with the three narrow components 
identified by Hines \etal~(1999).  
The results indicate that the continuum and broad \hb~are similarly 
polarized at 16\% $\pm$ 3\% and $PA$ = 98\arcdeg. 

Similar modeling at the Mg II line shows that the polarization of the 
broad Mg II line, at 29\%, is substantially higher than broad \hb, but is 
consistent with values seen in the imaging polarimetry. 
If real, it suggests that the intrinsic polarization rises to the blue, 
which is only possible if dust scattering is present.

Our fitting also shows that the core component of the narrow \oiii~lines 
has a small amount of polarization, 0.70\% $\pm$ 0.03\% at $PA$ = 90\arcdeg, 
which is different from that of the broad line and continuum. 
On the other hand, the \oii~\wave 3727 line is virtually unpolarized. 
We measure a polarization of 0.22\% $\pm$ 0.12\%. This indicates radial 
stratification of the NLR gas (see below).

\subsubsection{Stratification of Narrow-Line Polarization}

\begin{figure}
\plotone{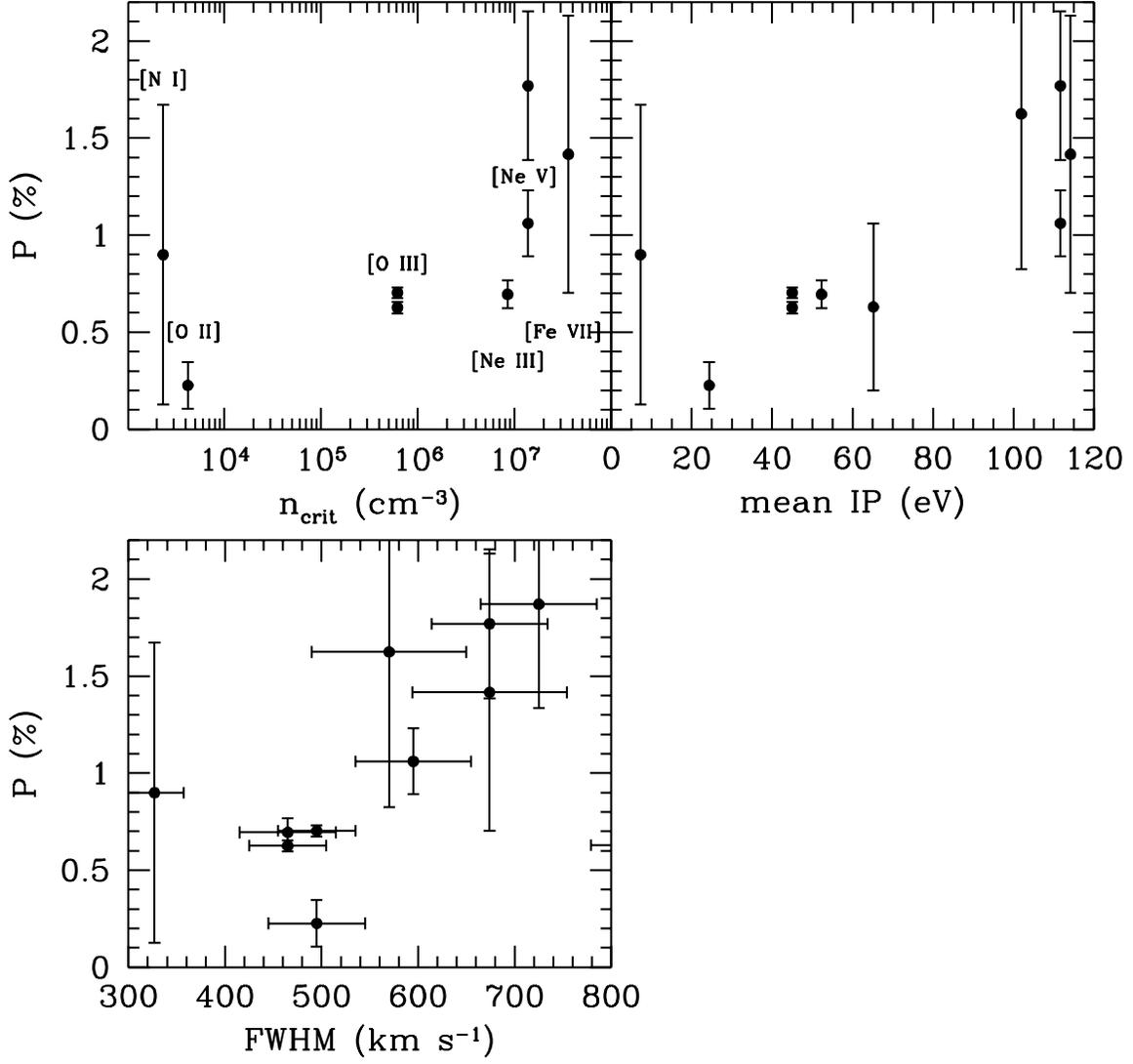}
\caption{Narrow forbidden line polarization as a function of $n_{crit}$, 
ionization potential (IP) and line widths. A clear trend of higher polarization
with higher $n_{crit}$, IP, and FWHM is evident.
\label{linep}}
\end{figure}

The high S/N data allowed us to measure the polarization of various 
emission lines in the nucleus. The results are in Table 2, where
we list the (biased) \p, \pa, their associated errors, the ionization 
potential (IP) of the lower and upper stages of ionization for the ions
of interest, and the critical density ($n_{crit}$) for collisional 
de-excitation of each line (calculated at 10$^4$ K).
The polarizations are flux-weighted means in the core component only, and
the uncertainties on \p~and \pa~are 1$\sigma$ statistical errors 
alone and do not include systematics such as those due to continuum placement
and data reduction.
In Figure \ref{linep}, we plot the measured line polarizations versus $n_{crit}$, 
the mean of the lower and upper IP, and FWHM. 
A clear trend is apparent in all three plots: 
lines of higher $n_{crit}$, IP and FWHM exhibit higher polarization. 
Such a correlation has also been observed in the LINER NGC 4258 
(\cite{barth99}). To our knowledge, NGC 4258 and \objn~are the
only two AGNs where this phenomenon has been shown to occur systematically
over a number of forbidden lines with a wide range of ionizations. 
The tendency for \oiii~to be more polarized than \oii, 
especially in radio galaxies (\cite{spell97}), has been known for some time, 
and an anisotropic \oiii~emission structure has been suspected as the 
cause (\cite{hes93}). 
Together with the well-known correlation of linewidth with $n_{crit}$ and IP
(e.g., \cite{dro86}) in Seyfert galaxies, 
this result provides strong indication that the narrow emission-line 
structure in \objn~is radially stratified, with lines of higher ionization 
and critical density being emitted closer to the nucleus, and therefore more 
likely to be partially occulted. The higher polarization results from  
less dilution by direct light.

\subsection{Physical Conditions and Diagnostic Diagrams}

The temperature determined from the \oiii~\wave 5007/\oiii~\wave 4363 
ratio (using Table 1) is $\approx$ 1.3$\times 10^4$ K for both NUC and EXT.
We also used the \nii~\wave 6583 line from the near-IR spectrum of 
Evans \etal~(1998) to measure the temperature from the 
\nii~\wave 5755/\wave 6583 ratio. 
This is only available for NUC, giving $T_e = 9000$ K. This is 
about 4000 K cooler than the \oiii~region, and may suggest that 
matter-bounded clouds are present (e.g., \cite{wbs97}).
The electron density estimated from the \ariv~\wave 4712/\wave 4740 ratio
is $\approx 10^4$ \cc~and 5,000 \cc~for NUC and EXT, respectively. 
This can be compared to a value of 6,000 \cc~for NUC determined from 
[Cl III] \wave 5518/\wave 5538 line ratio.
The [Ar IV] \waves 4712/\wave 4740 ratio is clearly better determined than 
the [Cl III] \wave 5518/\wave 5538 ratio since the stronger Ar lines are less
affected by starlight subtraction and yield more certain flux measurements.

To help us understand the rich emission-line spectrum of \objn, we 
compare the line ratios with model calculations.
In particular, we would like to understand the source of the ionization
and the excitation mechanism of a large number of
emission lines encompassing a wide range of ionization, 
from the low-excitation [N I] \wave 5200, and [S II] to the coronal lines 
such as \nev~and \fevii. 
Theoretical models from three main mechanisms are considered:

1) Simple photoionization of essentially radiation-bounded NLR 
clouds by the central source having a powerlaw radiation spectrum.
The emergent emission-line 
spectrum is characterized by the slope $\alpha$ of the ionizing continuum, 
and the ionization parameter $U=Q/4 \pi r^2 c n_e$, where $Q$ is the 
number of ionizing photons, $r$ is the radius and $n_e$ is the electron 
density.  
We build the AGN models with the photoionization code MAPPING Ic 
(\cite{ferr97}), using a power-law with index $\alpha = -1.5$.
The models are isobaric, so the density varies across the nebula. 
Recently, a new set of models has been computed for the NLR in an attempt
to account for more realistic physical conditions. In these models 
the line emission 
originates from a wide range of temperatures and densities, although the
observed emission-line spectrum results only from clouds best able to emit
it. These are called ``locally optimally emitting clouds'' (LOC)
models (Ferguson \etal~1997a, 1997b). 
We shall discuss these models but compare our observations only with the
simple, isobaric calculations that we made.

2) Two-component photoionization model of Binette \etal~(1996, 1997), 
consisting of both optically thin gas
(matter-bounded, MB) clouds and optically thick gas (ionization-bounded, IB)
clouds. The sequence is parameterized by the parameter \magp~which
is the ratio of the solid angle subtended by MB clouds relative to that of 
IB clouds, as viewed by the observer.
Most of the higher ionization lines (i.e., coronal lines and [O III])  
are emitted by MB clouds, which lie closer to the nucleus, filtering
the radiation field that reaches the IB clouds further away, where 
essentially all of the lower ionization lines (e.g, [N II], [O I]) originate 
(\cite{bin96}; \cite{bin97}).
A large value of \magp~indicates a larger weight to the MB component, and
therefore a higher excitation spectrum. 

3) Shocks and shock + precursors models of Dopita \& Sutherland (1995, 1996).
Mechanical energy deposited by velocity shocks due to turbulent cloud motions
or the outflowing radio jets can produce powerful local UV radiation which can 
ionize the gas. Two types of shock models have been considered: pure shocks,
in which mechanical energy from shocks is solely responsible for 
collisionally heating 
the gas, and the hybrid shocks + precursors, also known 
as ``photoionizing shocks'', in which the shock itself generates an abundant 
supply of ionizing photons in the hot post-shock gas and produce an extensive 
precursor H II region.

\subsubsection{Comparison to Model Predictions}

In Figures \ref{shock}--\ref{ibmb}, we present several diagnostic diagrams 
involving a number of emission lines with a wide range of ionization 
as an attempt to identify the main underlying
mechanism for ionizing the NLR gas in NUC and EXT. 
Model calculations based on the above three mechanisms are shown for 
comparison with the observations.
Simple power-law $U$ sequences are plotted for
density $n_e = 10^4$ \cc~(of the front layer) as determined from the 
nuclear gas.
Sequences with $n_e = 5000$ \cc~(as determined from the extended gas) 
produce very similar results and we do
not plot them on the diagrams. 
Two different sequences are shown for the IB+MB photoionization models of
Binette \etal~(1997), each characterized by the value of the ionization 
parameter {\it at the irradiated face of the MB component}, $U_o$:
high $U_o=0.5$ and low $U_o=0.05$.
The ionization parameter of the IB component is $6.5\times 10^{-4}$ for both. 
On all figures, the parameter $A_{M/I}$ goes from 0.01 to 100, 
increasing from left to right.
Pure shocks and shocks + precursors models are plotted as grids 
parameterized by the shock velocity $V$ and the magnetic parameter 
$B/\sqrt n$. $V$ varies from 150 to 500 \kms, and $B/\sqrt n$ varies from 
1 to 4 $\mu$G cm$^{-3/2}$.
All models are shown for solar abundances. 

The symbols have the following meanings: star denotes EXT and open circle
represents the total line flux from both the core and blue-shifted components
in NUC. The solid and open triangles indicate the core and blue-shifted 
emission line components of NUC, respectively. 
We shall now discuss these diagrams.

\subsubsection{Shocks and AGN photoionization}

Figure \ref{shock} shows that in two of the best optical diagrams for shock 
diagnostics, (\oiii~\wave 5007/\wave 4363 vs. He II \wave 4686/\hb~proposed 
by \cite{clk97}, \cite{vm99}, and [Ne V]$\lambda$3426/[Ne III]$\lambda$3869 vs. 
\oiii~\wave 5007/\hb~proposed by \cite{adt98}),
the positions of the NUC component lie well outside the range of model grids 
predicted by pure shocks and shocks + precursors. 
The shock + precursor models are a little more 
consistent with the points for the EXT gas.  
These diagrams clearly rejects pure shock as the dominant emission line 
mechanism in both the nuclear and extra-nuclear regions of \objn, 
as it tends to predict too low \oiii~\wave 5007. On the other hand, ionization-bounded
AGN photoionization models have the opposite problem, failing to agree with
the observed data by predicting too high \oiii~\wave 5007.

The presence of dust (see \S 3.3, 4.1) in the EXT also weakens considerably 
any arguments that shocks play a dominant role in the main energy input 
mechanism. If shocks were present the dust grains would likely be destroyed 
(Donahue \& Voit 1993; de Young 1999). 
Another argument against shocks is that the radio and emission line ``cone''
in \objn~are {\it misaligned} by as much as 38\arcdeg~(Hines \etal~1999), 
making shocks induced by the nuclear jets unlikely. 
Also, the \oiii~\wave 5007/\oii~\wave 3727 line ratio (a good indicator of 
the ionization parameter $U$) is observed to be 5.4 for NUC and 2.11 
for EXT. Thus $U$ does not go up or remain constant in the extra-nuclear 
regions, as would be the case if shocks + precursors provide extra local 
heating (e.g., as in 3C 299, \cite{fei99}), but actually drops, 
as expected from geometric dilution of the radiation field. 
Finally, the velocity widths of the emission lines in the extension are 
essentially unresolved $<$ 550 \kms, indicating that perturbations in 
the gas velocity by shocks, if any, are small.

In conclusion, we can rule out pure shocks as the ionizing mechanism. 
Shock + precursors are also ruled out in the nuclear region, 
but it is more ambiguous whether they exist in the extension.
Crawford \& Vanderriest (1996) have suggested that 
shock heating may increasingly become important in the extended gas. 
However, we do not find sufficient evidence in our data 
to support a dominant role by shocks even in the extensions.
More conclusive resolution between shocks and photoionization may come from 
additional diagnostics using UV lines (\cite{dop97}; \cite{adt98}), which 
are more sensitive to shocks than optical lines.

\subsubsection{The Need for Matter-Bounded Clouds}

\begin{figure}
\plotone{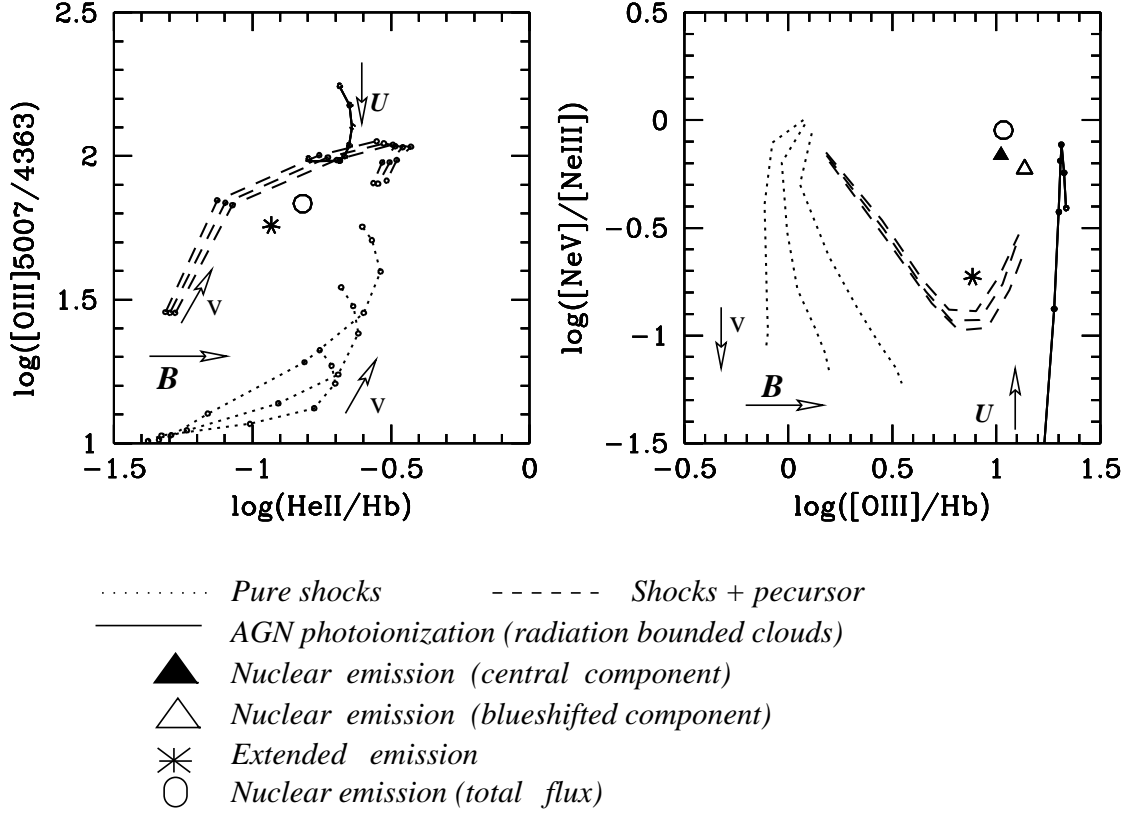}
\caption{Predictions of shock (dotted lines), shock+precursor (dashed lines) 
and AGN photoionization models with radiation bounded clouds and density 
$n_e=10^4$ \cc~(solid lines). $U$ is the ionization parameter
that defines the AGN sequence (see text). $U$ varies between 0.001 and 0.46.
Each shock (and shock+precursor) sequence is defined by a fixed value of the
magnetic parameter $\bf B$ ({$\bf B$} = $B/\sqrt n$) and a range of shock velocity
$v$ (150--500 \kms). Three shock sequences are shown, corresponding to 
$\bf B$ = 1--4 $\mu$G cm$^{-3/2}.$ These diagrams clearly reject pure shocks 
(radiative cooling of shocked gas) as the dominant emission line mechanism in the 
nuclear and extended gas. 
\label{shock}}
\end{figure}

As Figure \ref{shock}~indicates,
the models for the high-density IB photoionization sequence
predict too high \oiii~\wave 5007, typically giving \oiii/\hb \simgt~16. 
This has been a well-known problem for IB photoionization models, known 
as the ``temperature problem'':
the models predict too low a \oiii \wave 4363/\wave 5007 ratio and thus too 
low electronic temperatures (i.e., \cite{bin96}).
One attractive possibility that may solve this discrepancy is with the 
addition of MB clouds.
As already mentioned, the presence of MB clouds is suggested by the higher
temperature in the \oiii~emitting region compared to that in the 
\nii~emitting region in the nucleus, since the MB clouds, where 
most of the \oiii~originates, are expected to lie closer to the ionizing 
source (\cite{bin96}; \cite{wbs97}).

\begin{figure}
\includegraphics{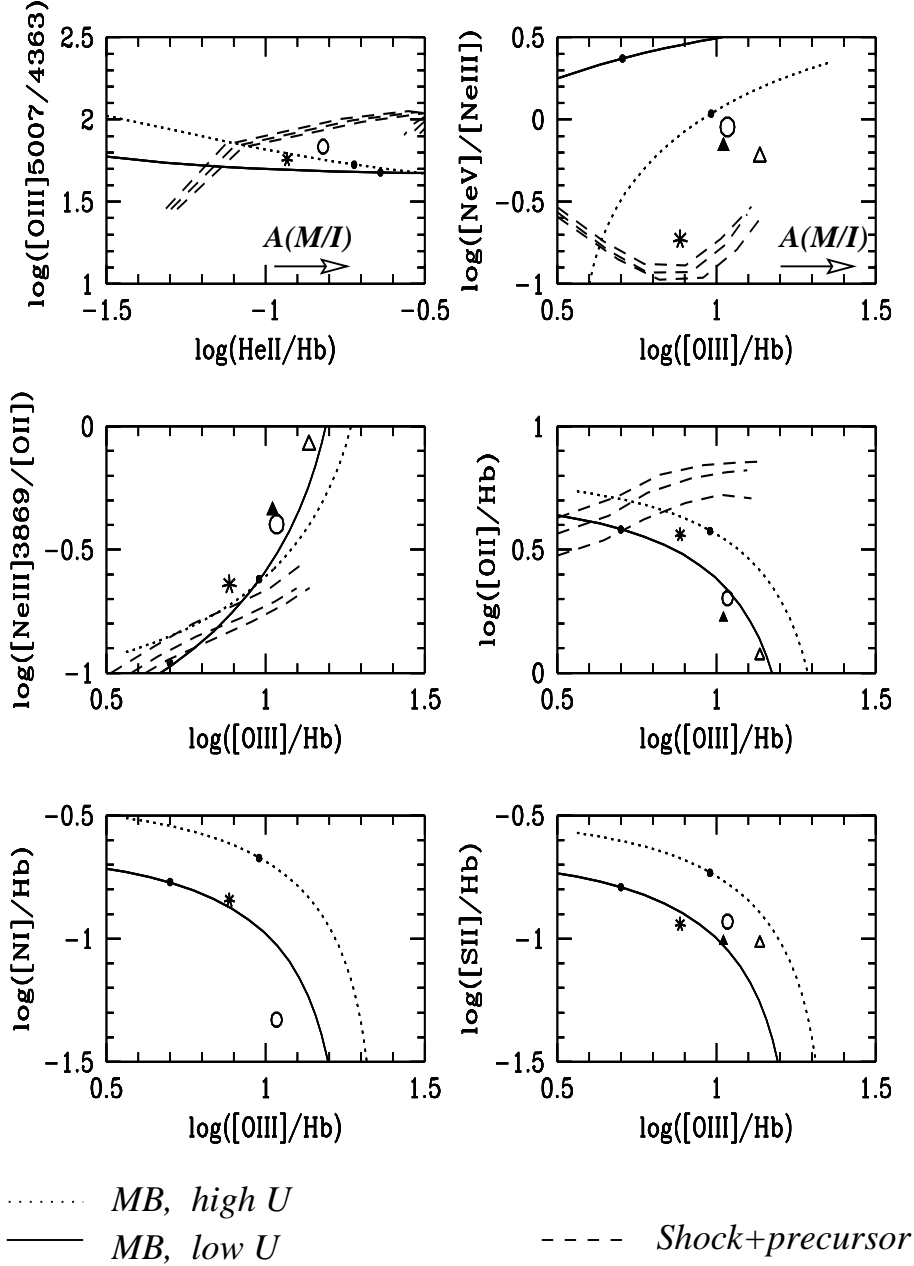}
\vspace{7in}
\caption{Shock+precursor model and AGN photoionization model (with
contribution from matter bounded clouds) predictions. The top two diagrams are
the same as in Figure \ref{shock}.
The shock+precursor models are plotted only on those diagrams where the
calculated line ratios are available. Symbols are as in
Figure \ref{shock}. The filled circles in the MB sequences represent the 
models with $A_{M/I}=1$ and the arrow in the upper panels shows increasing
$A_{M/I}$. The models with matter and ionization bounded clouds
are in good agreement with most line ratios except [Ne V]/[Ne III], whose 
predicted values are too high compared with the measurements. 
These diagrams reject shock+precursor models for the nuclear emission, where, 
therefore, AGN photoionization dominates. The discrimination
between AGN photoionization and shock+precursor models is not possible for the
extended gas. \label{ibmb}}
\end{figure}

In Figure \ref{ibmb}, we present additional diagnostic line ratios with comparison to 
the IB + MB as well as shock + precursor models.
The top two panels are the same as Figure \ref{shock}.  
As can be seen, the shock + precursor models clearly fail to fit all the 
NUC data points and miss the EXT points in most diagrams. 
In general, the addition of MB clouds is able to resolve the two main
difficulties of pure IB photoionization models: lines ratios of He II/\hb~and 
\oiii~\wave 5007/\wave 4363. However, these sequences also have problems.
For example, a common problem for all photoionization models is that [Ne V] is 
overpredicted (Fig. \ref{ibmb}, top right). 
Note also that only models with $A_{M/I}<1$ are able to fit the 
observed He II, but they have difficulty reproducing the high \oiii~\wave 5007/\hb, 
especially the low $U_o$ sequence (solid curve).
The high $U_o$ sequence (dotted curve in the plots) is consistent with {\it both}
the observed He II and \oiii~ratios, but it cannot reproduce the low-ionization lines
[O II]/\hb, [N I]/\hb~and [S II]/\hb, which are predicted to be too strong.
The low $U_o$ sequence has the opposite problem, agreeing well with these low-ionization
lines but failing to explain higher ionization lines like \nev~and \fevii, as well as
the weakness of He II for the observed [O III]/\hb.
Since it is the contribution from MB clouds that makes it possible to solve the
He II and temperature problems, and since the low-ionization lines are mostly due to
the IB component, a higher ionization parameter for the IB component might be consistent
with all data (recall that \cite{bin96} used a fixed $U$[IB]).

In summary, Figure \ref{ibmb} shows that shocks + precursors models fails to 
reproduce the observed data points in NUC and, to a lesser degree, in EXT.
Photoionization models with contribution from matter-bounded clouds are able to
reproduced the data well, although no single photoionization sequence is perfect. 
 
\section{Discussion}
\subsection{Dust in the Extended Emission-Line Gas}

In \S 3.3, we presented polarimetric evidence for dust in the scattering 
regions. Diagnostic line ratios also show evidence for dust in the emission 
line gas.
  
\subsubsection{High-Ionization and Fe lines} 

One of the most notable features of the emission line spectra of \objn~is  
the conspicuous absence of all species of Fe 
in EXT and their strong presence in NUC (see Fig. \ref{nucextsp} and Table 1). 
No other atomic species other than iron is seen to exhibit this behavior. 
This dramatic difference in Fe line strength between NUC and EXT
cannot be due to the lack of high ionization photons in EXT, 
for we see clearly in EXT strong high-ionization lines of [Ne V], which has
comparable ionization potential (97.1 eV) to [Fe VII] (100.0 eV).
Furthermore, lower ionization Fe lines like [Fe VI], [Fe V]
and perhaps [Fe III] are all completely absent from EXT. 
Coronal lines can be expected to be seen in the tenuous off-nuclear
extended gas (\cite{fer97b}), and indeed \nev~is seen in our EXT spectrum, 
and [Fe IX] has been observed in the extended emission line regions of
Circinus (\cite{oli94}; \cite{moor96}), Cygnus A (\cite{ogl97}) and
Tololo 0109$-$383 (Murayama, Taniguchi, \& Iwasawa 1998). 

The peak range of the line strength distribution function in the LOC models
of Ferguson \etal~for \nev~and \fevii~overlap substantially with each other. 
Therefore, under ``ordinary'' NLR conditions (i.e., solar abundances, free of 
dust) we would expect to see \fevii~wherever \nev~is seen. 
If \fevii~is as strong relative to other lines as observed in the nucleus, 
there should be no problem detecting it in EXT. 
In NUC, \fevii~\wave 6087 is about 3 times weaker than \nev~\wave 3246 
and 5.6 weaker than \hb. At this level, \fevii~should be even stronger than 
[N I] \wave 5200, or about as strong as He I \wave 4471, both of which are 
clearly seen in EXT. Even if it is 6--13 times weaker, as predicted by 
models of Binette \etal~(1997) or Ferguson \etal~(1997), it should be at 
least about as strong as [Ar IV] \waves 4712, 4740 or He I \wave 5876.
Somehow, virtually all gaseous Fe is removed from the extension gas.
With the inclusion of dust grains, \fevii~is expected to be about 74 times
weaker than \nev~(\cite{fer97b}). 
At this level, it would be impossible to detect with our data.
This strongly suggests that there is significant amount of dust in the 
extension allowing gas-phase Fe to deplete onto grains. 
This notion has been invoked to explain the emission-line spectra of 
Seyfert galaxies (Kingdon, Ferland, \& Feibelman 1995).
Neon is an inert gas and not expected to be depleted 
(see model calculations by \cite{fer97b}).
\fevii~\wave 6087 is a coronal line, and both 
Ferguson \etal~(1997b) and Binette \etal~(1997) have suggested that
coronal lines are not efficiently emitted in gas with dust. 
The lack of \fevii~\wave 6078 in EXT therefore is consistent with the presence
of dust. 

It is unlikely that the absence of Fe lines in EXT is  
due to the lower metallicity of the EXT region, for other metals
such as oxygen and nitrogen seem to be well represented. 
We would expect Fe to show a normal abundance as there exists a well-known
[Fe/H] to [O/H] relation for the nebula gas observed in the solar 
neighborhood and galactic bulge (e.g., Maciel 1999).

\subsubsection{Narrow Mg II}

Another element which may lend itself well to diagnostics for dust is Mg
since it is about as sensitive as Fe to depletion.
Unlike the higher-ionization forbidden Fe lines, \mgii~is complicated by 
several factors. 
It is a resonance line so it can be strongly affected by intervening dust 
extinction and gas absorption. 
On the other hand, like other low-ionization lines such as \sii, \oii, \nii~it may 
even get strengthened by photoelectronic heating of the gas by dust grains
(\cite{fer97a}). 

\begin{figure}
\plotone{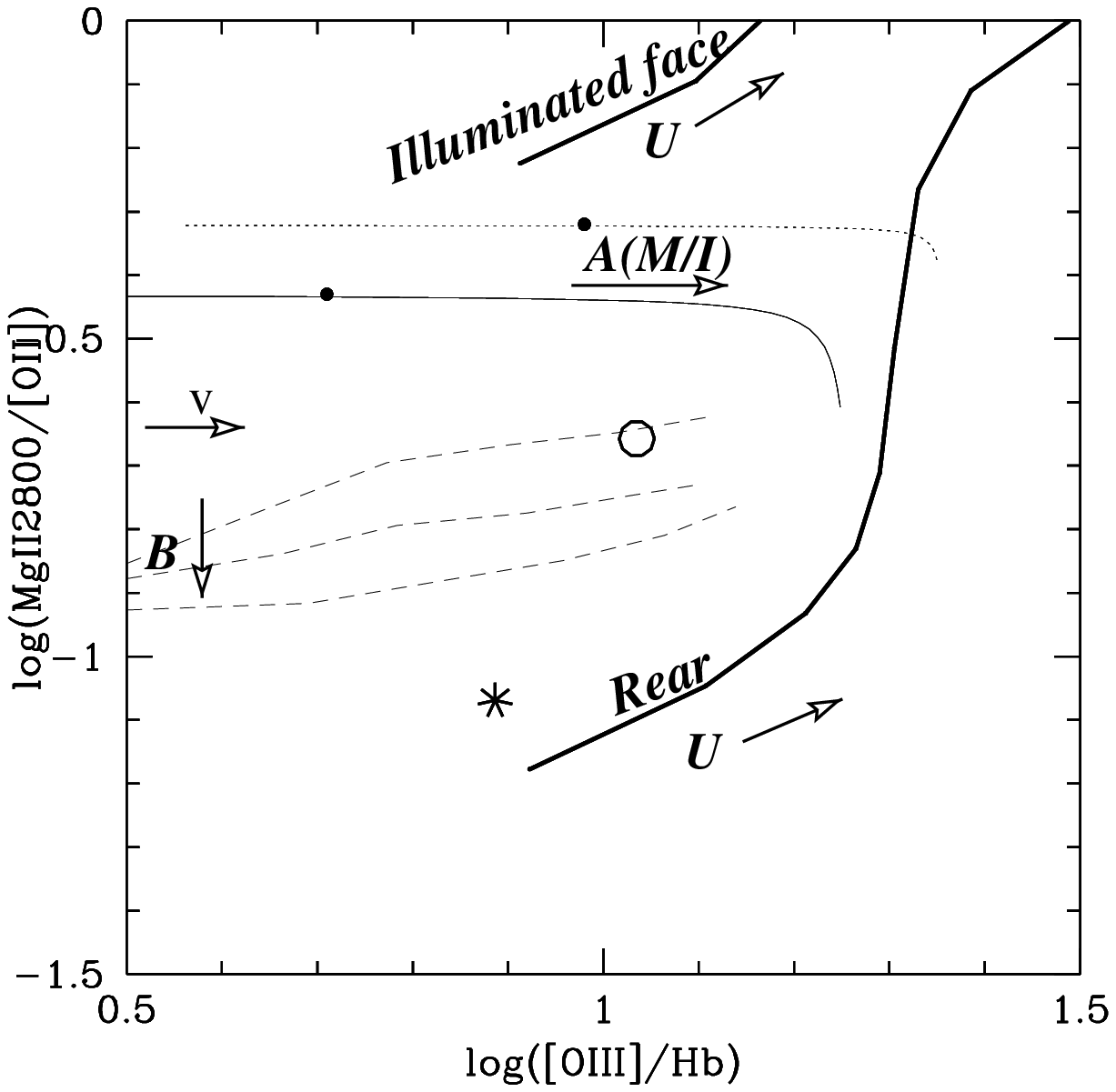}
\caption{Mg II \wave 2800/\oii~\wave 3727 vs. \oiii~\wave 5007/\hb.
The thick solid lines show the
photoionization $U$ sequence with radiation bounded clouds as shown in
Figure \ref{shock}, but taking into account perspective effects. 
The upper sequence
corresponds to the clouds seen from the illuminated face and the lower
sequence corresponds to the clouds seen from the rear. In this scenario,
only the extreme
case such that we are inside the ionization cone and see the clouds
from the rear could explain the weakness of Mg II $\lambda$2800.
See text for details. For comparison, shock + precursor and MB photoionization
models are also shown. 
Symbols and lines are as in Figure \ref{ibmb}. \label{mgii}}
\end{figure}

With these caveats, we now consider whether the resonant character of
Mg II \wave 2800 and its strong sensitivity to dust and gas absorption
suggest the presence of dust as well.
Figure \ref{mgii} shows the diagram of Mg II \wave 2800/\oii~\wave 3727 vs. 
\oiii~\wave 5007/\hb. 
The MB + IB photoionization models predict Mg II to be stronger than observed 
in both the nuclear and extended gas.
Shock + precursor models show reasonable agreement to the data, but they fail
for most other lines, and have been eliminated for the nuclear emission 
(see \S 3.4.2, 3.4.3).
One way to weaken Mg II \wave 2800 without dust is with
intervening absorbing gas, containing Mg II. Since Mg II is in ionized 
form, the most likely explanation is that the absorbing gas is in the 
ionized cones. We compare the predictions of the photoionization
models taking into account viewing effects: depending on the
viewing angle of the observer relative to the clouds, the spectrum of
the resonant lines, like Mg II \wave 2800 will be different (Villar-Martin,
Binette \& Fosbury 1996). 
If the clouds are viewed
from the illuminated face, the Mg II emission will be relatively stronger
than when seen from behind. In the former case, the photons
find a lower column density of Mg II (since the gas is highly ionized
towards the illuminated face) and they can escape the nebula freely. However,
photons that try to leave the cloud through the ``non-illuminated''
face of the cloud will encounter a much larger column density of Mg II and 
will be repeatedly absorbed. 

In Figure \ref{mgii} we show this effect. The upper thick solid line
shows the line ratios for a nebula viewed from the
illuminated surface. The lower thick solid line shows the line ratios for a nebula 
seen from behind. As can be seen, the difference is large. 
These models show that we need to view
the clouds directly from the back (the non-illuminated side) to reduce 
the Mg II emission to the low observed values. 
However, this extreme view is probably not the case, for if the clouds are 
seen from
behind, the observer would be inside the ionization cone and therefore, would
have a full view of a quasar or broad line radio galaxy.
Another possibility is that the absorbing gas is outside the
ionized cones, but in this case the ionization of the gas cannot be explained.

Thus again, the most reasonable explanation is that the emitting gas contains 
dust. It could be due to depletion (Mg is about as sensitive to depletion
as Fe), or to a combination of resonant scattering and dust absorption. 
Pure resonant scattering alone cannot explain the weakness of the line.
This conclusion is consistent with the indication for dust from the 
absence of Fe lines, as well as the polarized flux spectrum, which is
blue compared to a composite quasar spectrum (\S 3.3.2).

\subsection{Origin of the Extended Gas}
 
While the indication for dust in the nucleus and extended emission regions 
of \objn~comes as no surprise from one of the most luminous infrared 
emitters, this confirmation shows that the existence of dust is extensive 
over a large scale, and could have important implications for the origin of 
the extended gas. 

As suggested by Donahue \& Voit (1993) for the intracluster nebular gas 
that they studied, the presence of dust inferred in the extension implies 
that EXT cannot be gas condensed directly out of the X-ray emitting 
intra-cluster medium (ICM). 
This is because dust simply cannot survive in the hot ICM in any 
appreciable amount.
Any dust introduced in the hot ICM is expected to be 
destroyed long before the gas is cooled enough to be 
visible or gets transported to the cluster center by the cooling flow.
Villar-Martin \& Binette (1997) reached a similar conclusion in their
finding of dust in the extended emission-line regions of radio galaxies. 

Since there is no evidence that the metal abundances are anomalously 
low (\S 4.1.1), the most likely candidate for the extended gas is perhaps 
shreds of unfortunate members of the cluster being cannibalized by \objn. 
Active star forming regions, either in the galaxies themselves or triggered
by the interactions, are where copious amounts of dust are found.
This in turn implies that the extranuclear gas is unlikely to
be material expelled or outflowing from the nucleus of the central quasar.
$HST$ WFPC2 imaging of \objn~by Armus \etal~(1999) does indeed show what seem like 
whiskers and filaments, suggestive of stripped remains from nearby cluster 
members. 
In addition, recent $HST$ WFPC2 and NICMOS imaging studies of ULIRGs by Surace \etal~(1998)
and Scoville \etal~(2000) have shown that many galaxies contain numerous 
circumnuclear clusters of bright young stars (age $\sim$ 10$^7$--10$^8$ yrs), whose forming
regions are often associated with dust.

The external origin for the gas is supported by
the fact that many ULIRGs show evidence for
interaction or merging (e.g., Duc, Mirabel \& Maza 1997; Borne et al. 2000;
\cite{sco00}), which has been suggested to stimulate the ULIRG phase.
Velocity structure of the extended nebula of \objn~derived from integral field
spectroscopy by Crawford \& Vanderriest (1996) also suggests that the
gas is probably external in origin. However, these authors do not favor
the picture of the gas being merged with the cD galaxy, based on the low
relative velocity ($\sim$ 100--200 \kms) with respect to the nucleus.

\subsection{Scattering Geometry}

The polarizations of ULIRGs with hidden broad-line regions are generally
very high ($\sim$~20\%--30\%), suggesting that the inclination angle could be
fairly substantial (\simgt 35\arcdeg).
Under the assumption that the lack of a well-defined V-shape polarization 
fan suggests that the inclination angle $i$ does not exceed the half-opening
cone angle $\theta_c$ by a large amount,
we can use the morphology of the polarization image (Fig. \ref{polmap}) to 
constrain the scattering geometry of \objn, 
using the model of Brown \& McLean (1977) 
(see also \cite{mg90}; \cite{mgm91}; \cite{wil92}; 
\cite{bk93}; \cite{bro98}). 
Analysis by Hines \etal~(1999) has constrained the viewing inclination and 
half opening cone angle lie in the range $34\arcdeg < i < 41\arcdeg$ and
$15\arcdeg < \theta_c < 33\arcdeg$, based on \p = 20\%. 
With our higher polarization of $\sim$ 30\% from the broad
Mg II and imaging polarimetry, these limits are still consistent, but
we can eliminate the combination of both upper limits or lower limits for
$i$ and $\theta_c$ (i.e., $i=41\arcdeg$ $\theta_c = 33\arcdeg$, or $i=34\arcdeg$ $\theta_c = 15\arcdeg$), 
as they give too low \p. The same applies for the combination 
$i=34\arcdeg$, $\theta_c = 33\arcdeg$, in 
which case our line of sight is just outside the cone, but \p~is only 15\%.
\p~agrees better with observations if $i$ and $\theta_c$ are 
closer to the upper and lower limits, respectively, but in this case their
difference $i-\theta_c = 29\arcdeg$ is too large. If we limit the half opening
cone angle to be \simlt 40\arcdeg~constrained by the $HST$ image of 
Hines \etal~and Armus \etal~(1999), the difference $i-\theta_c$ is 
minimized to \simlt~10\arcdeg~for $i \sim 50\arcdeg$ and $\theta_c \sim 40\arcdeg$. 
Therefore, this appears to be the best scattering geometry for 
\objn~consistent with observations.

\section{Summary and Conclusions}

Our high-quality Keck spectropolarimetric and imaging polarimetric data 
confirm the high polarization and broad Balmer emission lines in the polarized 
flux spectra of \objn. In addition, we also detect broad Mg II in both total 
and polarized flux spectra of the nuclear and extension regions, indicating 
the presence of a hidden quasar visible only in scattered light. 
The narrow-line polarizations exhibit a strong positive correlation with 
line width, critical density and ionization potential of the transition,
indicating that the line emission arise from a radially stratified gas.
The lack of a clear fan-like morphology in the polarization image suggests
that our viewing angle ($i~\approx$~50\arcdeg) is not far outside the 
ionization cone ($\theta_c~\approx$~40\arcdeg) of \objn.

The emission-line spectra of \objn~are consistent with
AGN photoionization with contribution from both ionization-bounded and
matter-bounded clouds.
Under the assumption that photoionization by the central hidden quasar 
is the main operating mechanism, \fevii~should be detected in
the extended gas and Mg II should be much stronger than observed. 
The absence of all Fe lines
and weakness of Mg II in the extension can be readily explained if dust is 
mixed with the ionized gas of the NLR.
Furthermore, the fact that both the nuclear and the extended continua are
highly polarized with relatively blue polarized flux spectra compared to that
of the average quasar also supports the existence of dust, which could 
efficiently scatter and bluen the light.
The other possibility is that the extended gas simply is highly deficient 
in iron, or is metal poor. This does not seem very likely, since the 
strengths of the other metal lines appear to be in good agreement with solar 
abundance. Since grains remain intact, this in turns implies that shocks 
are probably not important in the extra-nuclear regions. 
Although the diagnostic diagrams offer somewhat inconclusive assessment 
of the shocks+precursor models for the extension region, there is no strong 
evidence for shocks in \objn~from the line ratios. Ionization by pure shocks
can be ruled out by the data. 
The existence of dust and lack of strong shocks in the extended gas also 
suggest that it is probably dismembered remnants of a cluster neighbor 
being disrupted or in the process of merging with the central cD galaxy. 

Two predictions follow directly from our interpretation of the current data:
The near-IR line [Fe II] 1.257 $\mu$m has been observed in the nucleus of 
\objn~by Soifer \etal~(1996), but should be absent from the extension, 
confirming the absence of gaseous Fe, and hence the lack of shocks and presence
of dust there. 
Likewise, [Ca II] \waves 7291, 7324 should be absent in EXT spectrum, 
since Ca is more than 10 times more sensitive to grain depletion than 
Fe or Mg (\cite{fer97b}).

\acknowledgments

We thank P. Ogle and A. Putney for assistance in the observations, and 
L. Binette for the use of his code Mapping Ic. We are grateful to the referee
for several comments, which have improved the clarity of the paper. 
The W. M. Keck Observatory is operated as a scientific partnership between
the California Institute of Technology and the University of California, 
made possible by the generous financial support of the W. M. Keck Foundation. 
This research has made use of the NASA/IPAC Extragalactic
Database (NED) which is operated by the Jet Propulsion Laboratory,
California Institute of Technology, under contract with the National
Aeronautics and Space Administration.

\newpage

\newpage
\begin{deluxetable}{llllllll}
\small
\tablecolumns{8}
\tablewidth{0pc}
\tablecaption{Emission Line Flux Ratios, Equivalent Widths and FWHMs}
\tablehead{
\colhead{} & \multicolumn{3}{c}{IRAS P09104+4109 Nucleus$^a$} & \multicolumn{1}{c}{} & \multicolumn{3} {c}{IRAS P09104+4109 Extension} \\
\cline{2-4} \cline{6-8} \\
\colhead{Line} &
\colhead{Flux Ratio$^b$} & \colhead{EW$^c$} & \colhead{FWHM$^d$} & \colhead{} &
\colhead{Flux Ratio$^{b}$} & \colhead{EW$^c$} & \colhead{FWHM$^d$}
} 
\startdata
He II \wave 2733             &  0.0275  &    2.7  &  712   &&          &         &         \nl 
[Mg V] \wave 2783            &  0.07183  &    7.0  &  1449  &&	       &         &         \nl     
Mg IIn \wave 2798            &  0.2584   &   25.0  &  1470  &&   0.3110 &    17.4 &  840    \nl
Mg IIb \wave 2798            &  0.3579   &   36.   &  14189 &&   0.71053&    55.6 &  14800  \nl 
He I \wave 2830 + O III \wave 2836  &  0.08345  &  8.1  &   &&   0.0324 &    1.84 &     \nl   
[Ar IV] \wave 2854           &  0.00638 &    0.52 &	    &&   0.0416 &    2.74 &       \nl   
[Ar IV] \wave 2868           &  0.00308 &    0.27 &	    &&          &         &       \nl   
[Mg V] \wave 2928            &  0.0176  &    1.8  &  1481  &&	       &         &       \nl          
He I \wave 2946              &  0.00995  &    0.95 &  762   &&	       &         &       \nl          
O III \wave 3047             &  0.0355  &    3.45 &	    &&          &         &       \nl
O III \wave 3133             &  0.1217   &   11.74 &	    &&  0.0367 &    2.79 &       \nl 
He I \wave 3188              &  0.0434  &    4.2  &  1200  &&   0.0361 &     2.78&       \nl 
He II \wave 3203             &  0.05546  &    5.3  &  635   &&   0.06463&     5.09&       \nl 
O III \wave 3312             &  0.0200   &    2.0  &	    &&          &         &       \nl
[Ne V] \wave 3346            &  0.2054   &   20.0  & 811, 674 &&  0.07033 &    5.65 &       \nl 
[Ne V] \wave 3426            &  0.5339   &   53.1  & 811, 595 &&  0.1561  &   12.7  &       \nl 
O III \wave 3444             &  0.0406  &    4.03 &  1044  &&	       &         &       \nl       
[Fe VII] \wave 3588          &  0.0476  &    4.76 & 635, 635 &&          &         &       \nl
[O II] \wave 3727            &  1.556    &  185.0  & 762, 495 &&  3.646   &  377.   &       \nl 
H12 (+ H11)                  &  0.1185   &   15.6  &	    &&  0.0326  &    3.82 &       \nl 
H11                          &   ...     &    ...  &	    &&  0.0408  &    5.1  &       \nl 
H10                          &  0.0429   &    5.93 &	    &&  0.0588  &    8.08 &       \nl 
H9                           &  0.0667   &    9.25 &	    &&  0.0734  &   10.2  &       \nl 
[Ne III] \wave 3869          &  0.8060   &  121.7  & 725, 465 &&  0.8343  &  123.3  &       \nl   
H8 + He I \wave 3889         &  0.1807   &   27.9  & 600 &&  0.1918  &   29.6  &       \nl   
H$\epsilon$ \wave 3970 + [Ne III] \wave 3967 &  0.3824   &   62.3  & 628, 580 &&  0.4041 &  59.2 &   \nl  
He I \wave 4026              &  0.0112   &    1.8  &	    &&  0.0203  &    2.9  &       \nl   
[S II] \wave 4071            &  0.09695  &   15.9  & 712, 750 &&  0.1145  &   17.0  &       \nl 
H$\delta$ \wave 4102         &  0.2163   &   35.4  & 835, 465 &&  0.2705  &   41.1  &  \nl 
He I \wave 4144              &  0.00980 &    1.6  &	    &&  0.0134  &    2.05 &       \nl
[Fe V] \wave 4229            &  0.07158  &   12.0  & 918, 859 &&          &         &       \nl
H$\gamma$ \wave 4340         &  0.4127   &   70.8  & 731, 480 &&  0.4655  &   73.9  &       \nl 
[O III] \wave 4363           &  0.1588   &   27.1  & 567 &&  0.1343  &   22.0  &       \nl 
He I \wave 4471              &  0.0359  &    6.4  & 906, 433 &&  0.0417  &    6.40 &       \nl 
He II \wave 4686n            &  0.1518   &   27.6  &  400   &&   0.1169 &    17.4 &       \nl 
He II \wave 4686b            &  0.1861   &   34.0  &  5020  &&	       &         &       \nl       
[Ar IV] \wave 4712           &  0.0150  &    2.7  &        &&   0.0245 &     3.73&       \nl 
[Ar IV] \wave 4740           &  0.0193  &    3.5  &        &&   0.025  &     3.90&       \nl 
H$\beta$ \wave 4861          &  1.000    &  184.   & 660, 465 &&  1.000   &  140.0  &       \nl 
[O III] \wave 4959           &  3.794    & 1050.   & 687, 495 &&  2.574   &  369.5  &       \nl 
[O III] \wave 5007           & 11.18     & 3541.   & 700, 465 &&  7.701   & 1084.   &       \nl 
[Fe VI] \wave 5146           &  0.0318  &    6.36 &  567   &&	       &         &       \nl       
[Fe VII] \wave 5159          &  0.0376  &    8.0  &  560   &&	       &         &       \nl       
[Fe VI] \wave 5176           &  0.0289  &    5.82 &	    &&          &         &       \nl
[N I] \wave 5200             &  0.05011  &   11.0  &        &&   0.1436 &    24.9 &       \nl 
[Fe III + Fe VII + Fe II]$^e$  &  0.0378 &    8.3  &	    &&          &         &       \nl
[Ca V] \wave 5309            &  0.0361  &    8.0  &	    &&          &         &       \nl
[Fe II + Fe VI]$^f$          &  0.0225  &    5.0  &	    &&          &         &       \nl
[Fe III + Fe VI]$^g$         &  0.0436  &    9.96 &	    &&          &         &       \nl
[Fe VI] \wave 5485           &  0.0142  &    3.26 &	    &&          &         &       \nl
[Cl III] \wave 5518          &  0.0117   &    2.46 &	    &&          &         &       \nl
[Cl III] \wave 5538          &  0.0149   &    3.14 &	    &&          &         &       \nl
[Ca VII] + [Fe VI]$^h$       &  0.0209   &    4.72 &	    &&          &         &       \nl
[Fe VI] \wave 5677           &  0.0153  &    3.44 &	    &&          &         &       \nl
[Fe VII] \wave 5720          &  0.1142   &   26.5  &	    &&          &         &       \nl
[N II] \wave 5755            &  0.0137  &    3.73 &	    &&  0.0359  &    5.91 &       \nl
He I \wave 5876              &  0.1174   &   29.5  & 680, 510 &&  0.1028  &   17.7  &       \nl 
[Fe VII] \wave 6087          &  0.1805  &   46.3  & 774, 674 &&          &         &       \nl

\tablenotetext{a}{Flux ratios and EWs are measured from starlight-subtracted spectrum.}
\tablenotetext{b}{Observed line flux ratios relative to narrow H$\beta$. Values
for the nucleus are the sum of both the central and blueshifted components corrected for $E(B-V)=0.24$.}
\tablenotetext{c}{Rest-frame equivalent widths in \AA.}
\tablenotetext{d}{FWHMs (\kms) have been corrected for instrumental resolution 
of 550 \kms. The two values for the nucleus denote the central and blueshifted
component, respectively.}
\tablenotetext{e}{Probable identification: [Fe III] \wave 5270 + [Fe VII] \wave 5276 + [Fe II] \wave 5262.}
\tablenotetext{f}{Probable identification: [Fe II] \wave 5334 + [Fe VI] \wave 5335.}
\tablenotetext{g}{Probable identification: [Fe III] \wave 5412 + [Fe VI] \waves 5424, 5427.}
\tablenotetext{h}{Probable identification: [Ca VII] \wave 5615 + [Fe VI] \wave 5631.}

\enddata
\end{deluxetable}

\newpage

\begin{deluxetable}{lccccc}
\tablewidth{0pc}
\tablecaption{Forbidden Line Polarization}
\tablehead{
\colhead{Line} & \colhead{$P$ (\%)} & \colhead{PA (\arcdeg)} & \colhead{Lower IP (eV)} & \colhead{Upper IP (eV)} & \colhead{$n_{crit}$ (cm$^{-3}$)} }
\startdata
[Ne V] \wave 3346   & 1.77 $\pm$ 0.38 & 104.0 $\pm$ 6.2  & 97.1 & 126.2  &  $1.38 \times 10^7$ \nl
[Ne V] \wave 3426   & 1.06 $\pm$ 0.17 & 102.4 $\pm$ 4.6  & 97.1 & 126.2  &  $1.38 \times 10^7$ \nl
[O II] \wave 3727   & 0.23 $\pm$ 0.12 & 124.8 $\pm$ 15.2 & 13.6 &  35.1  &  $4.20 \times 10^3$ \nl
[Ne III] \wave 3869 & 0.70 $\pm$ 0.07 &  93.8 $\pm$ 3.0  & 41.0 &  63.5  &  $8.48 \times 10^6$ \nl
[Fe V] \wave 4229   & 0.63 $\pm$ 0.43 &  56.9 $\pm$ 17.7 & 54.8  & 75.5  & \nodata \nl
[O III] \wave 4959  & 0.70 $\pm$ 0.03 &  87.7 $\pm$ 1.1  & 35.1 &  54.9  &  $6.16 \times 10^5$ \nl
[O III] \wave 5007  & 0.63 $\pm$ 0.03 &  85.8 $\pm$ 1.3  & 35.1 &  54.9  &  $6.16 \times 10^5$ \nl
[N I] \wave 5200    & 0.90 $\pm$ 0.77 & 118.0 $\pm$ 24.6 & 0.0  &  14.5  &  $2.34 \times 10^3$ \nl
[Fe VII] \wave 6087 & 1.42 $\pm$ 0.71 &  68.4 $\pm$ 14.4 & 100.0 & 128.3 &  $3.6~ \times 10^7$  \nl

\enddata
\end{deluxetable}

\end{document}